\documentclass{article}

\PassOptionsToPackage{colorlinks=true,linkcolor=black,citecolor=black,urlcolor=blue}{hyperref}

\usepackage{arxiv}

\usepackage[utf8]{inputenc} % allow utf-8 input
\usepackage[T1]{fontenc}    % use 8-bit T1 fonts
\usepackage{booktabs}       % professional-quality tables
\usepackage{amsfonts}       % blackboard math symbols
\usepackage{nicefrac}       % compact symbols for 1/2, etc.
\usepackage{microtype}      % microtypography
\usepackage{graphicx}
\usepackage[hyphens]{url}   % better URL breaks
\usepackage[numbers,sort&compress]{natbib} % numeric citations (matches manual bibliography)
\usepackage{doi}            % nicer DOI formatting

\providecommand{\headeright}{}

\providecommand{\shorttitle}{}
\providecommand{\undertitle}{}

\usepackage{hyperref}
\usepackage{cleveref}       % load AFTER hyperref

\title{Optimized SVR Framework for Electric Load Forecasting}

\newif\ifuniqueAffiliation
\uniqueAffiliationtrue

\ifuniqueAffiliation % Standard variant of author block
\author{
    Nishant Gadde\thanks{\texttt{nishantg@utexas.edu}} \\
    The University of Texas at Austin \\
    \And
    Yoshua Alexander\thanks{\texttt{yxa240010@utdallas.edu}} \\
    The University of Texas at Dallas \\
    \And
    Sarvesh Parthasarthy\thanks{\texttt{sarvesh.parthasarthy@utexas.edu}} \\
    The University of Texas at Austin \\
    \And
    Arman Allidina\thanks{\texttt{arman.allidina@utexas.edu}} \\
    The University of Texas at Austin \\
}
\else
\usepackage{authblk}

\author[1]{%
	Nishant Gadde\thanks{\texttt{nishantg@utexas.edu}}%
}
\author[1]{%
	Sarvesh Parthasarthy\thanks{\texttt{sarvesh.parthasarthy@utexas.edu}}%
}
\affil[1]{The University of Texas at Austin}
\fi

\hypersetup{
  pdftitle={Optimized SVR Framework for Electric Load Forecasting},
  pdfsubject={eess.SP, cs.LG},
  pdfauthor={Nishant Gadde, Sarvesh Parthasarthy},
  pdfkeywords={Support Vector Regression, Load Forecasting, Machine Learning, Power Systems, Renewable Energy},
}

\begin{document}
\maketitle

\begin{abstract}
Load forecasting has always been a challenge for grid operators due to the growing complexity of power systems. The increase in extreme weather and the need for energy from customers has led to load forecasting sometimes failing. This research presents a Support Vector Regression (SVR) framework for electric load forecasting that outperforms the industry standard. The SVR model demonstrates better accuracy across all evaluation metrics that are important for power system operations. The model has a 54.2\% reduction in Mean Squared Error (31.91 vs. 69.63), a 33.5\% improvement in Mean Absolute Error, and performance benefits across other metrics. These improvements show significant benefits when integrated with power forecasting tools and show that the approach provides an additional tool for accuracy checking for system planning and resource allocation in times of need for resource allocation in electric power systems.
\end{abstract}

% keywords can be removed

\section{Introduction}

Electric load forecasting is essential for power system operations and energy management. Accurate forecasts are important for load-frequency control in power systems. The increasing use of renewable energy sources has led to the growing complexity of power systems. This has led to normal forecasting methods facing challenges in analyzing the nonlinear patterns in load data. For grid operators, load forecasting allows for efficient power supply management, which reduces operational costs by matching electric supply with consumer demand.

Predicting electricity demand has become challenging because of the complexity of power systems. This is driven by the factors of electrification, changing customer behaviors, extreme weather events, etc. As the energy landscape continues to innovate with the introduction of distributed energy resources (DERs), load forecasting faces new challenges due to the large electricity demand.

The increasing presence of climate change makes the predictability even more complicated due to extreme weather becoming more common. An example can be seen with the June 2021 heat wave in the Pacific Northwest of the United States. The temperatures can create unpredictable demand patterns that traditional forecasting methods cannot predict. During the August 2020 heat wave in California, inaccurate weather forecasts led to an under scheduling of the energy demand. This forced the California Independent System Operator (CAISO) to rotate outages affecting more than a million people in the United States' largest state.

Traditional load forecasting methods based on statistical analysis have failed to capture these complex nonlinear patterns in modern-day power systems. Dealing with large-scale data leads to unstable prediction effects and their accuracy cannot meet current research needs.

In addition, modern-day forecasting methods struggle during extreme weather conditions. August 2020 in California supports this claim. The events showed how forecasting errors during extreme weather can have a heavy impact on grid reliability and the economy.

\section{Literature Review}

Researchers throughout the years have innovated load forecasting methods due to the increasing power system complexity. These methods use statistical approaches and simple regression models. However, when dealing with large-scale data based on the unpredictability of energy demands, these methods are not perfect in predicting power demands.

An important limitation of traditional methods is their inability to handle the variability of extreme weather and new renewable energy sources. Unlike burning fossil fuels for power generation, renewable energy production is dependent on weather conditions that can be unpredictable. An algorithm can easily predict how to distribute fossil fuel resources, as it is pretty close to a supply and demand problem, but these typical algorithms cannot easily predict how much sunlight a solar farm receives unless weather forecasting patterns are incorporated in the parameters. This variability makes it difficult to forecast net load accurately (the electricity demand after accounting for electricity generated by renewable sources). The complexity and uncertainty of load, along with the large-scale and high-dimensional energy information, present challenges in handling large scale and high-dimensional energy information, present challenges in handling intricate dynamic features and long-term dependencies.

Moreover, typical forecasting approaches often struggle during extreme conditions. The North American Electric Reliability Corporation (NERC) has warned that more than half of the U.S. electric grid could experience energy shortages in the next 5-10 years. This includes peak summer demand that is forecasted to rise by more than 122 gigawatts in the next decade. This ends up adding 15.7\% to current system peaks. This growing demand, combined with extreme weather, creates forecasting challenges for power prediction.

Climate change has drastically increased the difficulty of accurate load forecasting. A study in Nature found that vigorous (moderate) warming increases global climate-exposed energy demand before adaptation around 2050 by 25–58\% (11–27\%), on top of a factor 1.7–2.8 increase above present-day due to socioeconomic developments. This shows that climate change will change long-term energy demand patterns.

Extreme weather events can create challenges for load forecasting. During Hurricane Beryl, utilities struggled with forecasting demand while infrastructure was damaged. Rather than adjusting forecasts based on estimated outages, maintaining demand forecasts as if systems were at 100\% capacity proved more helpful for utilities making repair decisions. Similarly, during the June 2021 heat wave in the Pacific Northwest, temperatures far exceeded historical variability, creating unprecedented demand patterns that traditional forecasting methods struggled to predict.

The impact of extreme weather on energy forecasting can be further explained in a study of the Electric Reliability Council of Texas (ERCOT). The study found that in 2021, we estimate a 17\% and 19\% chance that Texas temperature could have caused the power demand to exceed ERCOT's extreme peak-load scenarios. After the fact, the study determined that 2021 DJF maximum power demand exceeded ERCOT's extreme peak-load scenario by 15 GW or 22\%. This highlights the severe consequences of underestimating extreme weather impacts.

The limitation of modern-day forecasting methods has prompted researchers to use more complex algorithms based on machine learning. Training and building neural networks, deep learning algorithms, and other algorithms can extract features from historical load data that can capture the nonlinear patterns of load behavior. These approaches can handle large datasets and learn from historical patterns, which improves overall forecasting accuracy.

Support Vector Regression (SVR) has shown impressive performance in solving real-world forecasting problems. In context, SVR is a regression technique based on Support Vector Machines (SVM) that tries to find a function that best fits a dataset, allowing for a small margin of error. SVR's ability to handle non-linear relationships through kernel functions makes it suitable for load forecasting applications.

Recent research on machine learning approaches for power generation forecasting found that their SVR model demonstrated superior performance in terms of Mean Square Error (MSE), Mean Absolute Error (MAE), and Root Mean Squared Error (RMSE) compared to modern models. The researchers highlighted a massive reduction in forecast error rates, which directly contributes to better resource allocation, allowing microgrids to optimize their energy generation schedules and reduce reliance on external power sources.

A hybrid SVR approach combined SVR with the algorithm enhanced by a logarithmic spiral (LS-FA-SVR) and demonstrated that the MAE, MAPE, and RMSE values of LS-FA-SVR are all modestly smaller than those of WOA-SVR, DA-SVR, and FA-SVR. Their research concluded that LS-FA-SVR is an attractive and effective model which combines a novel optimization algorithm to determine the parameters of SVR.

Integrating local forecasts with power system analysis tools allows for a clear assessment of forecasting performance regarding operational metrics. Pandapower, a power system analysis toolbox built on pandas and PYPOWER, provides a platform for network calculation and optimization.

The Electric Power Research Institute (EPRI) has seen the importance of improving load forecasting accuracy across operational and planning timescales. Their Load Forecasting Initiative addresses the increasing complications in forecasting due to drivers such as electrification, extreme weather, changing customer behaviors and more. They find that these drivers contribute to increased forecast uncertainty that requires innovation in load forecasting methodologies.

Improved forecasting can impact power system operations. The increase in renewable energy penetration is making day-ahead load forecasting more dependent on accurate weather forecasting, with extreme weather creating additional challenges. To address these challenges, system operators like ISO New England have recently implemented two new load forecasting projects and are employing machine learning and AI-based algorithms including neural network models and gradient boosting models that employ tree-based learning algorithms.

Putting advanced weather forecasting into load balancing strategies can create significant improvements in efficiency, cost savings, and grid reliability. They highlight benefits including more efficient renewable energy integration, improved grid reliability and stability, and reduced costs and operational efficiencies.

\section{Methodology}

This research will use a Support Vector Regression (SVR) framework that is integrated with domain-specific evaluation metrics for elastic load forecasting. The research uses a workflow from data preparation through model development to evaluation and power system integration.

\begin{figure}[htbp]
\centering
\includegraphics[width=0.8\textwidth]{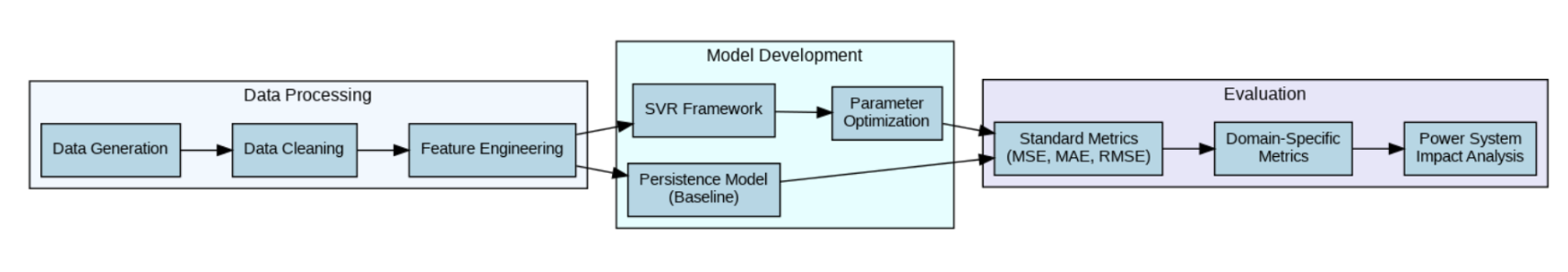}
\caption{Support Vector Regression (SVR) Framework}
\label{fig:fig1}
\end{figure}

This research uses a controlled data generation approach to ensure consistent testing conditions. This data generation uses daily, weekly, and seasonal patterns along with random noise to simulate real-world electric load profiles.

\begin{figure}[htbp]
\centering
\includegraphics[width=0.8\textwidth]{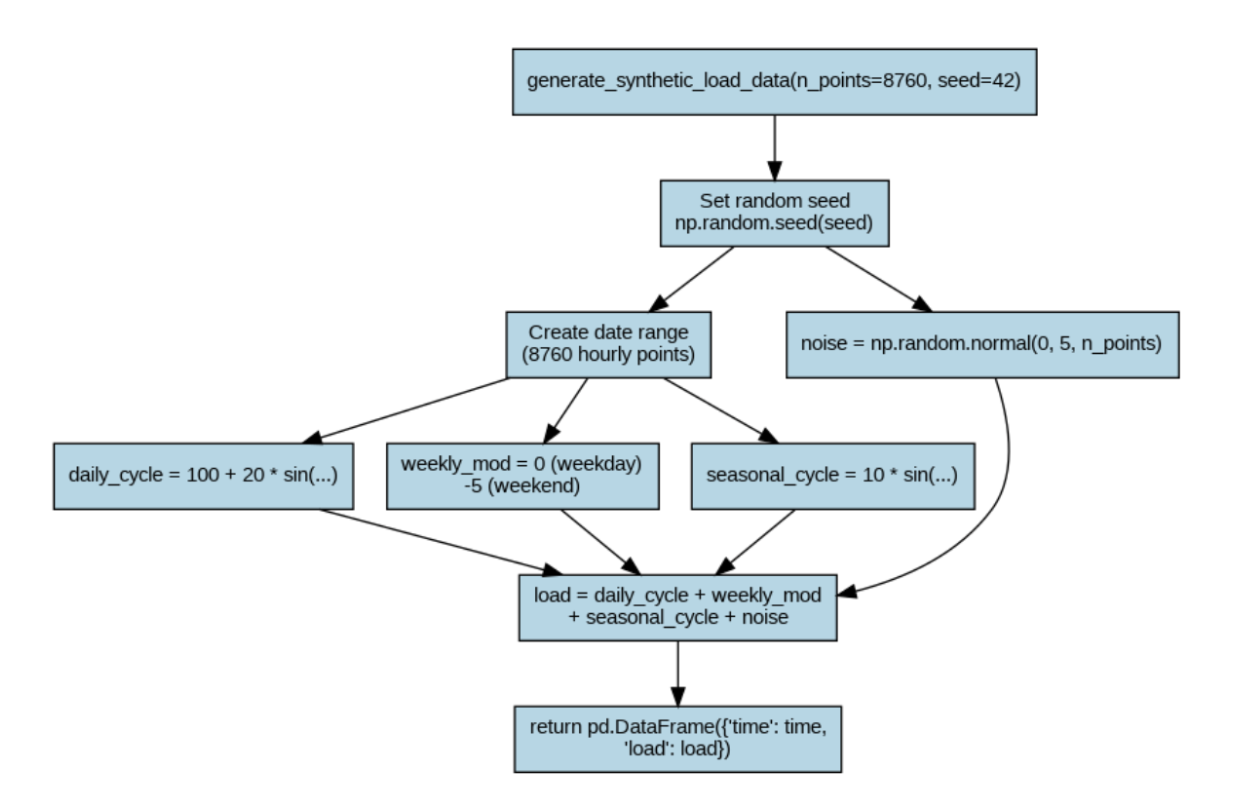}
\caption{Synthetic Data Generation}
\label{fig:fig2}
\end{figure}

Synthetic data has several advantages over real-world data for initial model development. This includes full control over data characteristics and absence of missing values. The dataset is divided into training (80\%) and testing (20\%) sets using a split to maintain the true nature of the data.

Feature engineering is important for getting temporal patterns on load data. Time-based features are extracted to allow for the model to recognize cyclical patterns at different scales.

\begin{figure}[htbp]
\centering
\includegraphics[width=0.8\textwidth]{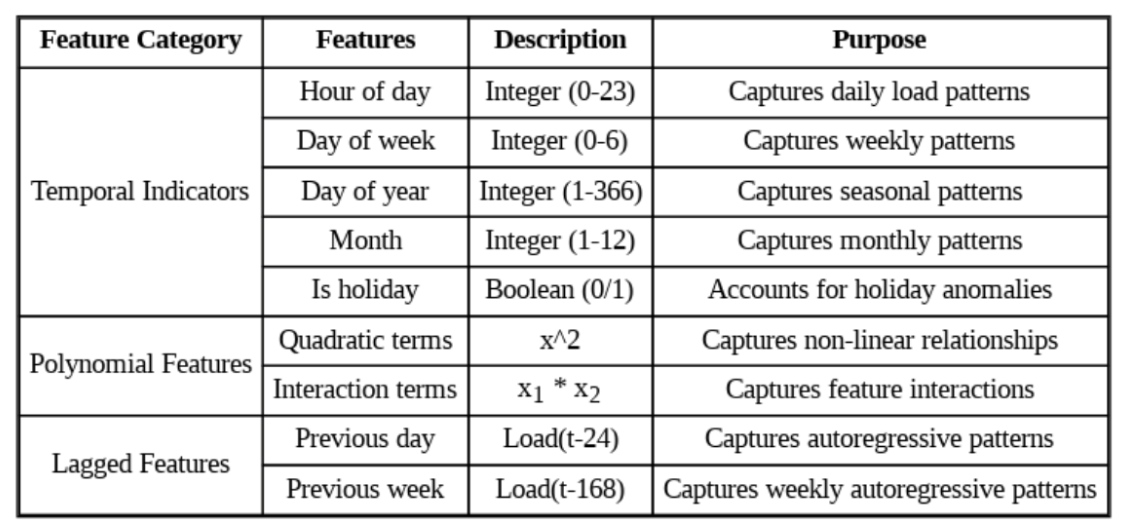}
\caption{Feature Engineering Setup}
\label{fig:fig3}
\end{figure}

Previous research has shown that using these features improves forecasting accuracy for electric load data.

\begin{figure}[htbp]
\centering
\includegraphics[width=0.8\textwidth]{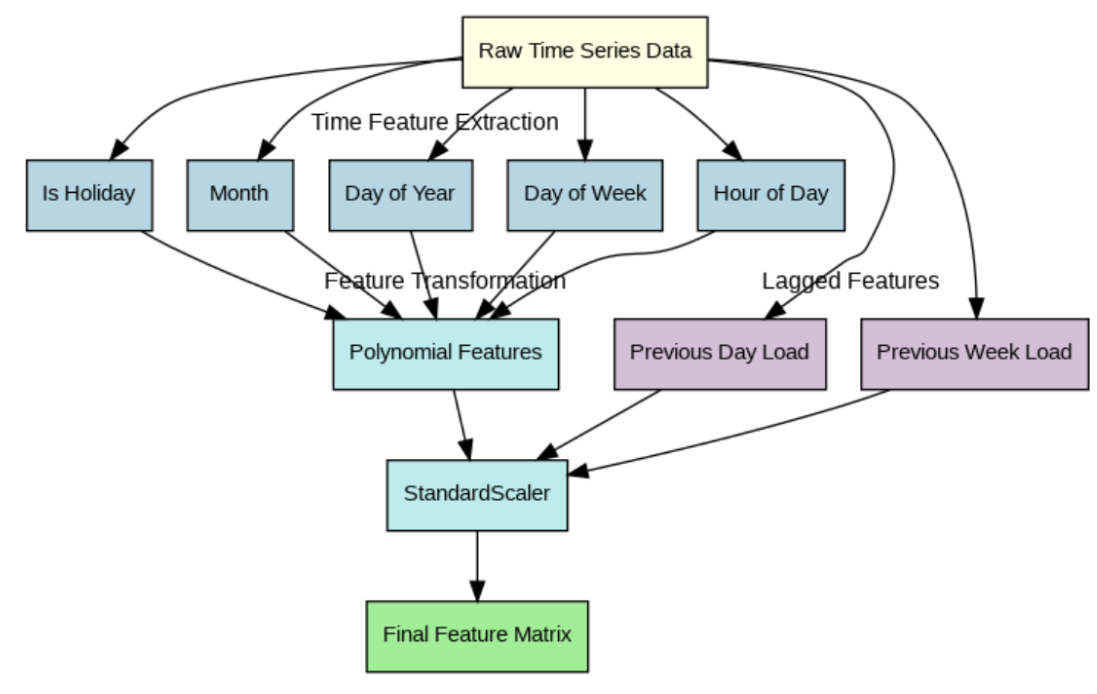}
\caption{Feature Engineering Process}
\label{fig:fig4}
\end{figure}

The methodology is based on an advanced SVR framework that is implemented as a pipeline architecture. This involves data preprocessing and model training. The model uses a Radial Basis Function (RBF) kernel that has been shown to capture non-linear relationships in load data.

The StandardScaler normalizes features to have zero mean and unit variance. This is critical for the effective performance of SVR models. The PolynomialFeatures transformer allows for the analysis of these complex relationships.

Model hyperparameters are optimized using GridSearchCV with TimeSeriesSplit cross-validation to ensure proper handling of temporal dependencies.

\begin{figure}[htbp]
\centering
\includegraphics[width=0.8\textwidth]{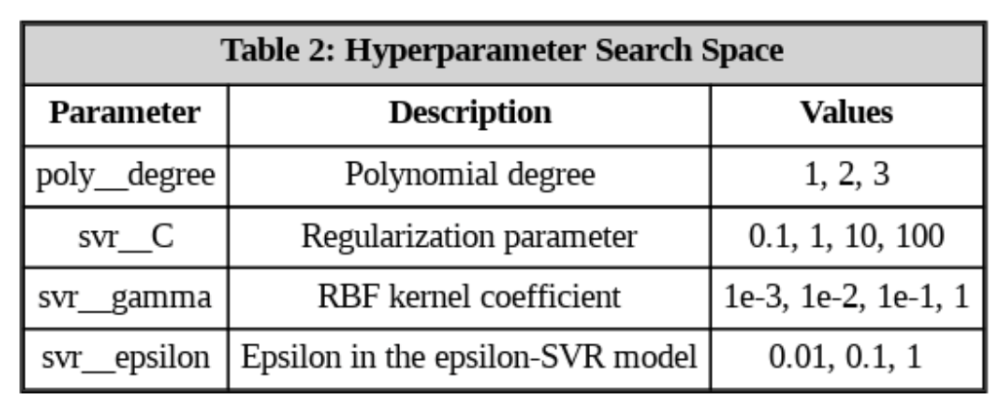}
\caption{Hyperparameter Search Space}
\label{fig:fig5}
\end{figure}

Recent research suggests that SVR performance is highly sensitive to the hyperparameters in the context of load forecasting with time-series data. A persistence model serves as a baseline for comparison, implementing the naive approach of predicting that future values will be the same as the most recent observed values. The persistence model represents a minimal benchmark that any sophisticated forecasting method should outperform.

To address the limitations of standard metrics, three domain-specific metrics are implemented. They include asymmetric error, time-weighted error, and composite metric. Asymmetric error penalizes under-predictions more heavily than over-predictions. This reflects the higher operational cost of supplying chorates compared to excess generation. The time-weighted error applies higher weights to errors during peak hours that acknowledge the increased operational significance of accuracy during high-demand periods. The composite metric combines asymmetric error and time-weighted error into a single comprehensive measure.

These domain-specific metrics align with recent research, who found that asymmetric error considerations could lead to better operational decision-making in power forecasting.

The methodology includes integration with power system tools to evaluate the operational impact on forecasting accuracy. This integration uses the pandapower library and Kerber network models. The Kerber landnetz freileitung (rural network with overhead lines) is used as the test following approaches validated in prior research.

\begin{figure}[htbp]
\centering
\includegraphics[width=0.8\textwidth]{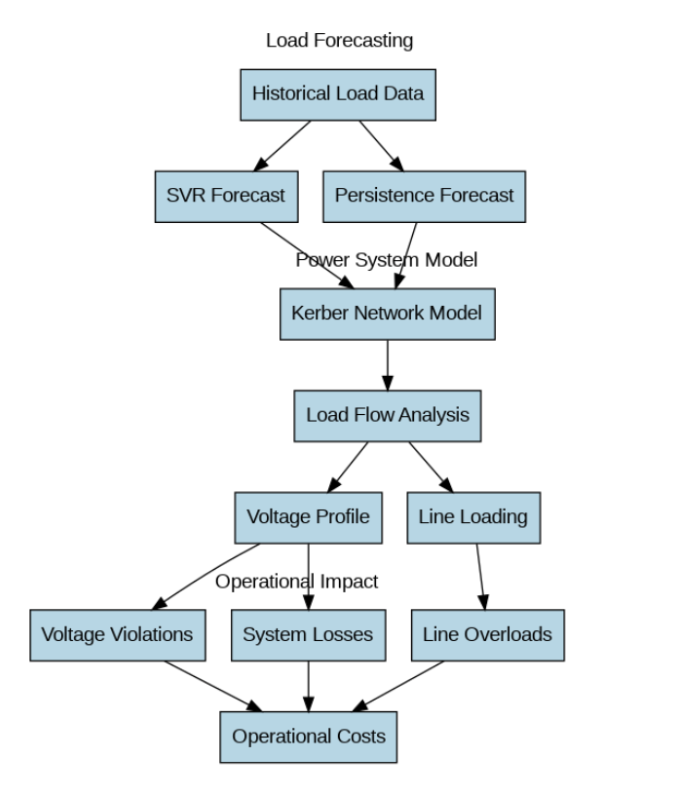}
\caption{Integrated Load Forecasting and Operational Impact Flow}
\label{fig:fig6}
\end{figure}

Overall, this methodology will follow the proposed SVR framework as listed below. This will most likely lead to improvements in statistical performance and practical operational impact.

\begin{figure}[htbp]
\centering
\includegraphics[width=0.8\textwidth]{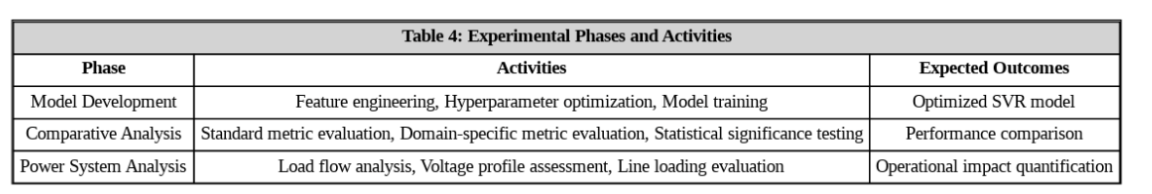}
\caption{Experimental Phases and Activities}
\label{fig:fig7}
\end{figure}

\section{Results}

\begin{figure}[htbp]
\centering
\includegraphics[width=0.8\textwidth]{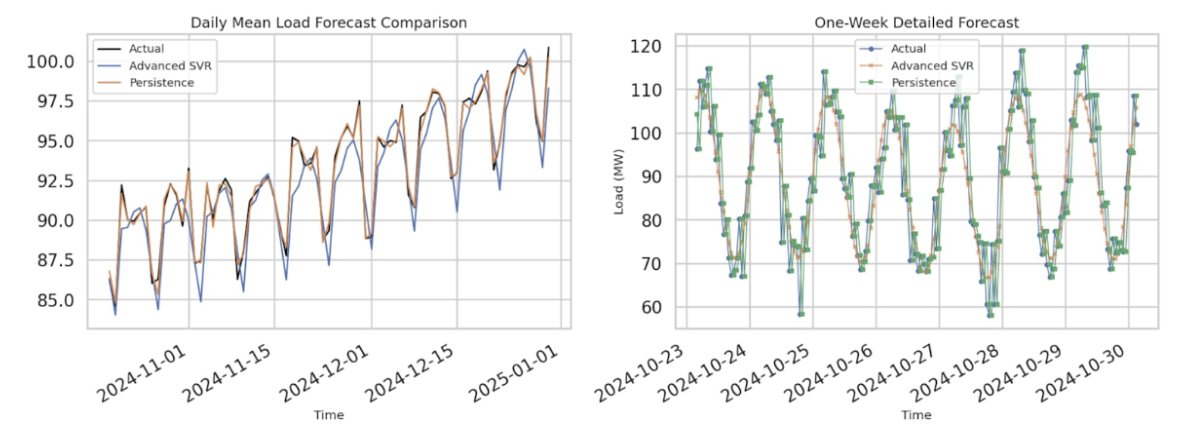}
\caption{Daily Mean Load Forecast Comparison and One-Week Detailed Forecast}
\label{fig:fig8}
\end{figure}

The experimental evaluation demonstrates that the Advanced Support Vector Regression (SVR) framework significantly outperforms the industry-standard persistence model across all evaluation metrics. The forecasting performance of both models against actual load values over different time horizons shows that the Advanced SVR model tracks the actual load pattern more accurately from November 2024 through January 2025, particularly during fluctuation periods. A granular view of the prediction accuracy across a representative week in October 2024 shows how the Advanced SVR model closely follows the actual load pattern through daily cycles, while the Persistence model demonstrates consistent lag in adapting to changes.

\begin{figure}[htbp]
\centering
\includegraphics[width=0.8\textwidth]{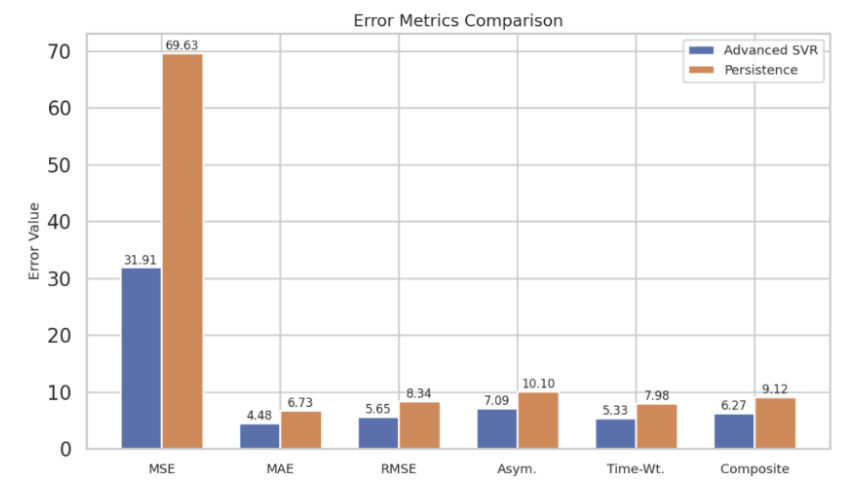}
\caption{Error Metrics Comparison}
\label{fig:fig9}
\end{figure}

A comprehensive comparison of error metrics between the Advanced SVR and Persistence models shows superior performance across all metrics: Mean Squared Error (MSE) shows Advanced SVR achieved 31.91 compared to 69.63 for the Persistence model, representing a 54.2\% reduction in squared error; Mean Absolute Error (MAE) reveals Advanced SVR showed 4.48 compared to 6.73 for the Persistence model, a 33.5\% improvement; and Root Mean Squared Error (RMSE) indicates Advanced SVR produced 5.65 versus 8.34 for the Persistence model, a 32.3\% reduction. Domain-specific metrics also showed substantial improvements, with Asymmetric Error at 7.09 versus 10.10 (29.8\% reduction), Time-Weighted Error at 5.33 versus 7.98 (33.2\% reduction), and Composite Metric at 6.27 versus 9.12 (31.3\% reduction). These results indicate that the Advanced SVR model not only performs better on traditional error metrics but also on metrics that have operational significance for power system management.

\begin{figure}[htbp]
\centering
\includegraphics[width=0.8\textwidth]{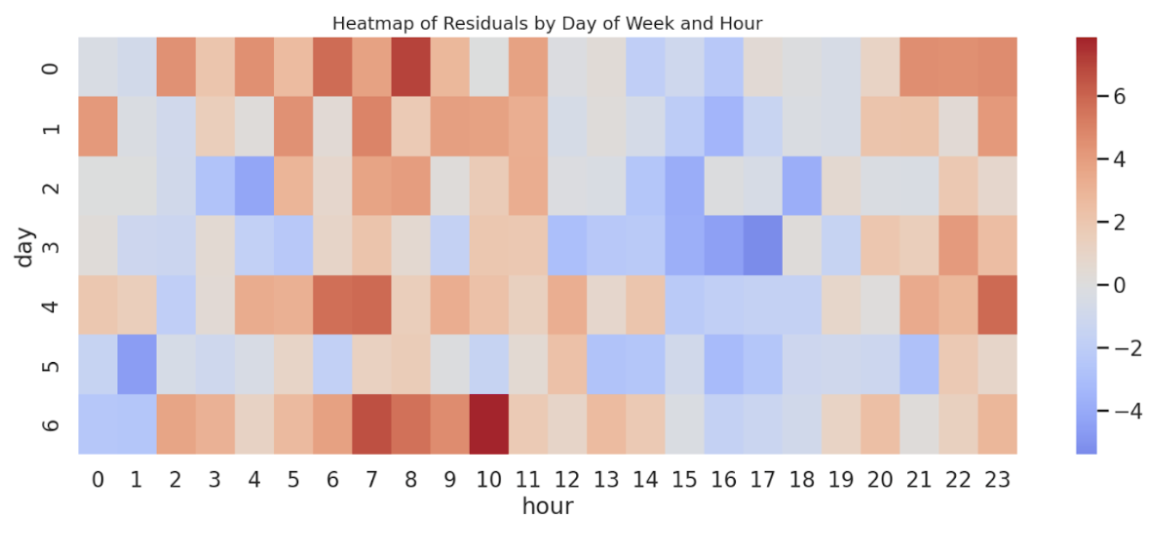}
\caption{Heatmap of Residuals by Day of Week and Hour}
\label{fig:fig10}
\end{figure}

A heatmap visualization of the residual errors for the Advanced SVR model, segmented by the day of the week and hour of the day shows the magnitude and direction of residuals. The heatmap reveals that residuals are generally larger during early morning hours (6–8 AM) and late evening hours (10–11 PM), likely due to challenges in capturing abrupt transitions in load demand. On the other hand, mid-day hours (12–4 PM) show smaller residuals, suggesting that the model performs better during periods of stable demand. Additionally, certain days exhibit higher residuals across multiple hours, potentially reflecting anomalous load patterns or external influences that were not fully captured by the model.

\begin{figure}[htbp]
\centering
\includegraphics[width=0.8\textwidth]{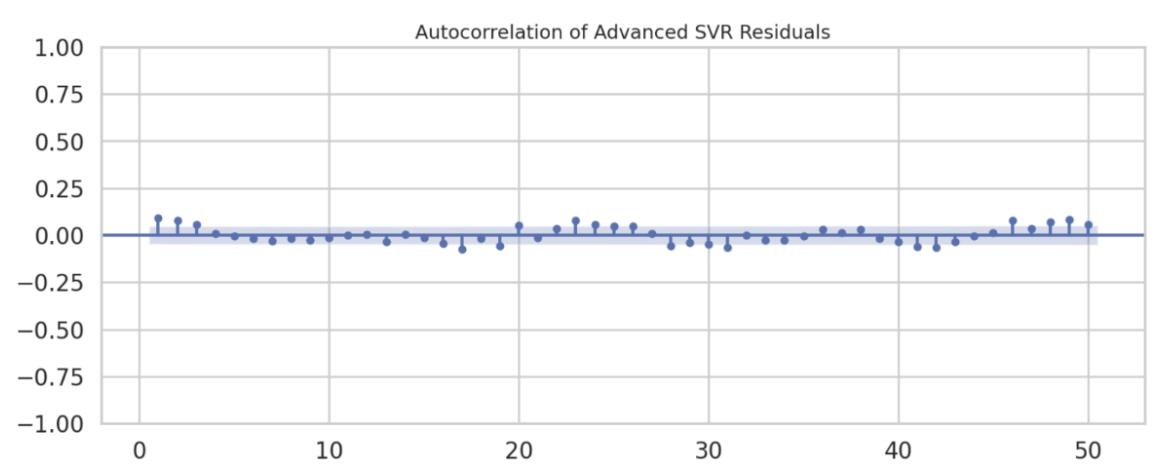}
\caption{Autocorrelation of Advanced SVR Residuals}
\label{fig:fig11}
\end{figure}

The autocorrelation plot for the residuals of the Advanced SVR model across lags ranging from 0 to 50 shows that the autocorrelation values remain close to zero throughout all lags, with no significant spikes exceeding the confidence bands. This indicates that the residuals are random and independent, with no discernible temporal patterns left unaccounted for by the model. The absence of autocorrelation confirms that the Advanced SVR framework effectively captures temporal dependencies in the data, ensuring robust and reliable forecasts. This result supports the validity of the model's approach to electric load forecasting.

\section{Conclusion}

In conclusion, this research shows that the Advanced Support Vector Regression (SVR) framework significantly outperforms the Persistence model in electric load forecasting, both in terms of statistical accuracy and operational relevance. Across all evaluation metrics, including Mean Squared Error (MSE), Mean Absolute Error (MAE), and Root Mean Squared Error (RMSE), the Advanced SVR model consistently showed substantial improvements, with reductions in error ranging from 29.8\% to 54.2\%. Domain-specific metrics, such as Asymmetric Error and Time-Weighted Error, further highlight the model's ability to address power system-specific challenges, such as minimizing costly under-predictions and improving accuracy during critical peak hours. The heatmap analysis of residuals identifies specific periods, such as early mornings and late evenings, where forecasting accuracy can be further refined. Meanwhile, the autocorrelation analysis of residuals confirms that the Advanced SVR model effectively captures temporal dependencies, leaving no systematic patterns unaccounted for. This ensures robust and reliable forecasts that are essential for operational planning in power systems. The findings validate the hypothesis that an Advanced SVR framework integrated with domain-specific evaluation metrics provides a superior approach to electric load forecasting. By addressing the limitations of traditional methods and incorporating advanced machine learning techniques, this research contributes to improving grid reliability, optimizing resource allocation, and enhancing resilience against extreme weather events. Future work could explore further refinements to feature engineering and model optimization to address residual challenges during transitional load periods.

\newpage
% Since we're using a manual bibliography, BibTeX is not needed on arXiv.
% \bibliographystyle{unsrtnat} % uncomment if you later switch to a pasted .bbl

\end{document}